\documentclass[aps,prd,onecolumn,groupedaddress,showpacs,nofootinbib,amssymb]{revtex4}
\usepackage[dvips]{graphicx}
\usepackage{amssymb}
\usepackage{amsmath}
\usepackage{graphicx}
\usepackage{amsfonts}
\usepackage{bm}
\usepackage{amsmath,amssymb,theorem}

\begin{document}

\title{Autonomous Dynamical System Approach for Inflationary Gauss-Bonnet Modified Gravity}
\author{V.K. Oikonomou,$^{1,2,3}$\,\thanks{v.k.oikonomou1979@gmail.com}}
\affiliation{$^{1)}$Department of Physics, Aristotle University of Thessaloniki, Thessaloniki 54124, Greece\\
$^{2)}$ Laboratory for Theoretical Cosmology, Tomsk State University
of Control Systems
and Radioelectronics, 634050 Tomsk, Russia (TUSUR)\\
$^{3)}$ Tomsk State Pedagogical University, 634061 Tomsk, Russia\\
}

\tolerance=5000

\begin{abstract}
In this paper we shall analyze the $f(\mathcal{G})$ gravity phase
space, in the case that the corresponding dynamical system is
autonomous. In order to make the dynamical system autonomous, we
shall appropriately choose the independent variables, and we shall
analyze the evolution of the variables numerically, emphasizing on
the inflationary attractors. As we demonstrate, the dynamical system
has only one de Sitter fixed point, which is unstable, with the
instability being traced in one of the independent variables. This
result holds true both in the presence and in the absence of matter
and radiation perfect fluids. We argue that this instability could
loosely be viewed as an indication of graceful exit in the
$f(\mathcal{G})$ theory of gravity.

 has
only
\end{abstract}

\pacs{95.35.+d, 98.80.-k, 98.80.Cq, 95.36.+x}

\maketitle

\section{Introduction}

The dark energy and the dark matter problem are two of the main
unanswered problems in modern theoretical cosmology. The dark energy
refers to the late-time acceleration that our Universe is currently
experiencing, which was observed in the late 90's,
\cite{Riess:1998cb} and it is unarguably the most intriguing problem
for current and future theoretical research. The dark matter problem
is bit older, and the most popular explanation comes from particle
physics, in the context of which, dark matter is materialized by a
non-interacting particle, see for example \cite{Oikonomou:2006mh},
for a deep analysis of various aspects of dark matter candidates.
Both the aforementioned problems find appealing theoretical
explanations in the context of modified gravity, in it's various
aspects, see the reviews \cite{reviews1,reviews2,reviews3,reviews4},
for extensive presentations on the subject. Also, a recently
proposed theory that may successfully mimic dark matter in a
geometric way, is offered by mimetic theories of gravity
\cite{Chamseddine:2013kea}, see \cite{Sebastiani:2016ras} for a
recent review.

Among the various modified gravity proposals that exist in the
literature, an appealing proposal is Gauss-Bonnet $f(\mathcal{G})$
gravity, where $\mathcal{G}$ is the Gauss-Bonnet invariant. The
latter, in four dimensions, is defined as $\mathcal{G}=R^2-4R_{\mu
\nu}R^{\mu \nu}+R_{\mu \nu \rho \sigma}R^{\mu \nu \rho \sigma}$, and
with $R_{\mu \nu}$, $R_{\mu \nu \rho \sigma}$ are the Ricci tensor
and the Riemann tensor respectively. From the mathematical form of
the Gauss-Bonnet invariant, it is obvious that such a theory could
contain higher derivatives, which would render the theory not so
appealing, since it would be hard to tackle with. However, although
this is true in principle, in the $f(\mathcal{G})$ case, the
resulting picture turns out that it is greatly simplified in four
dimensions, since the theory contains second order derivatives,
hence it can be studied more easily than initially one may think. In
the context of $f(\mathcal{G})$ it is possible to describe various
cosmological evolutions \cite{Li:2007jm}, like late-time
acceleration or the unification of inflation with late-time
acceleration
\cite{Nojiri:2005jg,Nojiri:2005am,Cognola:2006eg,Elizalde:2010jx,Izumi:2014loa,Oikonomou:2016rrv}.
Apart from the inflationary paradigm, several proposals exist for
the description of bouncing cosmologies in the context of modified
$f(\mathcal{G})$ gravity, see for example
\cite{Oikonomou:2015qha,Escofet:2015gpa,Makarenko:2017vuk,new1,new2}.

In this paper the focus is on the dynamical system corresponding to
$f(\mathcal{G})$ gravity. Particularly, we shall be interested in
providing an autonomous dynamical system analysis, as this was
performed in Ref. \cite{odintsovoikonomou}. In the literature there
exist various studies on the subject in the context of
Einstein-Hilbert or modified gravity, see for example
\cite{Boehmer:2014vea,Bohmer:2010re,Goheer:2007wu,Leon:2014yua,Guo:2013swa,Leon:2010pu,deSouza:2007zpn,Giacomini:2017yuk,Kofinas:2014aka,Leon:2012mt,Gonzalez:2006cj,Alho:2016gzi,Biswas:2015cva,Muller:2014qja,Mirza:2014nfa,Rippl:1995bg,Ivanov:2011vy,Khurshudyan:2016qox,Boko:2016mwr,Granda:2017dlx},
and also \cite{Odintsov:2015wwp}, in which an autonomous dynamical
system approach was also informally used. The question that
naturally springs to mind is why we should use an autonomous
dynamical system approach in the first place. The answer to this
comes from the fact that for a non-autonomous dynamical system,
finding the fixed points may not suffice or maybe one could be lead
to not correct conclusions, regarding the phase space. Particularly,
the stability of the dynamical system is not guaranteed by solely
using theorems like the Hartman-Grobman, which describe autonomous
dynamical systems. In order to further support our claim, let us
exemplify it by using a characteristic example, which can be found
in \cite{dynsystemsbook,odintsovoikonomou}: Consider the one
dimensional dynamical system $\dot{x}=-x+t$, the solution of which
is, $x(t)=t-1+e^{-t}(x_0+1)$. By looking the solution, it can easily
be seen that all the solutions (for the various initial conditions,
which have impact on $x_0$) tend to $t-1$ for $t\to \infty$.
However, if the standard fixed point analysis of non-autonomous
dynamical systems is applied in this case, it yields the result that
the only fixed point is $x=t$, which is not a solution to the
dynamical system. In addition, by using standard non-autonomous
approaches, it can be found that the vector field is deflected from
the solution $x(t)=t-1$, which is not correct as we discussed
earlier. Hence, in many cases, the non-autonomous dynamical systems
analysis may not suffice, so the autonomous dynamical system
analysis is compelling.

In view of the above, the purpose of this paper is to provide a
dynamical system analysis of the $f(\mathcal{G})$ modified gravity
theory, focusing on inflationary attractors. We shall provide the
general form of the $f(\mathcal{G})$ gravity autonomous dynamical
system, in the presence of cold matter and radiation, and we shall
analyze in detail the dynamical system, focusing on inflationary
attractors. The time dependence of the resulting dynamical system is
solely contained in the parameter $m$, which is equal to
$m=-\frac{\ddot{H}}{H^3}$, where $H$ is the Hubble rate. Hence, we
shall assume that this parameter takes constant-values, and
therefore the dynamical system turns out to be autonomous. We shall
be particularly interested in the case $m=0$, which corresponds to
the de Sitter vacuum. We shall find the fixed points of the
autonomous dynamical system and we shall examine the stability. Also
we shall investigate numerically the evolution of the dynamical
variables, in terms of the $e$-foldings number, focusing on values
in the range  $N=[0,60]$, which are more interesting when
inflationary theories are considered. As we demonstrate, in all the
cases, an unstable de Sitter attractor exists, which may reflect the
possibility of having an intrinsic mechanism for the graceful exit
in $f(\mathcal{G})$ theories.

Before we start our presentation, let us briefly present the
geometric framework we shall use, which will be a flat
Friedmann-Robertson-Walker (FRW) metric, with line element,
\begin{equation}\label{frw}
ds^2 = - dt^2 + a(t)^2 \sum_{i=1,2,3} \left(dx^i\right)^2\, ,
\end{equation}
with $a(t)$ being the scale factor.

\section{Autonomous Dynamical System of the $f(G)$ Gravity}

In this section we shall present in brief some basic features of
$f(\mathcal{G})$ gravity, and we shall investigate how to construct
an autonomous dynamical system by using the cosmological equations
and some appropriately chosen variables.

The $f(\mathcal{G})$ gravitational action is equal to
\cite{Nojiri:2005jg,Nojiri:2005am,Cognola:2006eg,Elizalde:2010jx,Izumi:2014loa}:
\begin{equation}
\label{GB1b} S=\int d^4x\sqrt{-g} \left(\frac{1}{2\kappa^2}R +
f(\mathcal{G}) + \mathcal{L}_\mathrm{matter}\right)\, ,
\end{equation}
and upon variation with respect to the metric tensor  $g_{\mu\nu}$,
the gravitational equations of motion are,
\begin{align}
\label{GB4b}& 0= \frac{1}{2\kappa^2}\left(- R^{\mu\nu} + \frac{1}{2}
g^{\mu\nu} R\right) + T_\mathrm{matter}^{\mu\nu} +
\frac{1}{2}g^{\mu\nu} f(\mathcal{G}) -2 f'(\mathcal{G}) R R^{\mu\nu}
\\ \notag &  + 4f'(\mathcal{G})R^\mu_{\ \rho} R^{\nu\rho} -2 f'(\mathcal{G})
R^{\mu\rho\sigma\tau} R^\nu_{\ \rho\sigma\tau} - 4 f'(\mathcal{G})
R^{\mu\rho\sigma\nu}R_{\rho\sigma} + 2 \left( \nabla^\mu \nabla^\nu
f'(\mathcal{G})\right)R \\ \notag & - 2 g^{\mu\nu} \left( \nabla^2
f'(\mathcal{G})\right) R - 4 \left( \nabla_\rho \nabla^\mu
f'(\mathcal{G})\right) R^{\nu\rho} - 4 \left( \nabla_\rho \nabla^\nu
f'(\mathcal{G})\right)R^{\mu\rho} \\ \notag &  + 4 \left( \nabla^2
f'(\mathcal{G}) \right)R^{\mu\nu} + 4g^{\mu\nu} \left( \nabla_{\rho}
\nabla_\sigma f'(\mathcal{G}) \right) R^{\rho\sigma}
 - 4 \left(\nabla_\rho \nabla_\sigma f'(\mathcal{G}) \right)
R^{\mu\rho\nu\sigma} \, .
\end{align}
It is notable that the equations of motion Eq.~(\ref{GB4b}) do not
contain any higher derivative terms. For the flat FRW metric of Eq.
(\ref{frw}), the gravitational equations take the following form,
\begin{align}
\label{eqnsfggrav} & 6H^2+f(\mathcal{G})-\mathcal{G}f'(\mathcal{G})
+24H^3\dot{\mathcal{G}}f''(\mathcal{G})+\rho_m+\rho_r=0 \\ \notag &
-2\dot{H}=-8H^3\dot{F}+16H\dot{H}\dot{F}+8H^2\ddot{F}+\frac{4}{3}\rho_r+\rho_m
\, ,
\end{align}
where the ``dot'' denotes differentiation with respect to the cosmic
time, $rho_m$ and $\rho_r$ stand for the energy density of the
baryonic matter and radiation respectively. Finally, $F$ in Eq.
(\ref{eqnsfggrav}) stands for,
\begin{equation}\label{generaldefinitonoff}
F(\mathcal{G})=\frac{\partial f(\mathcal{G})}{\partial
\mathcal{G}}\, .
\end{equation}
Also, for the flat FRW Universe of Eq. (\ref{frw}), the Gauss-Bonnet
invariant $\mathcal{G}$, takes the following form,
\begin{equation}\label{gaussbonnetinvariant}
\mathcal{G} = 24 \left( H^2 \dot H + H^4 \right)\, .
\end{equation}
By using the cosmological equations (\ref{eqnsfggrav}), we shall
construct a dynamical system that can be rendered autonomous, if a
set of appropriately chosen variables is used, and to this end we
introduce the following variables,
\begin{equation}\label{variablesslowdown}
x_1=-\frac{\dot{F}(\mathcal{G})}{F(\mathcal{G})H},\,\,\,x_2=-\frac{f(\mathcal{G})}{3H^2},\,\,\,x_3=
\frac{\mathcal{G}}{24H^4},\,\,\,x_4=\frac{\rho_r}{3H^2},\,\,\,x_5=\frac{\rho_m}{3H^2},\,\,\,
x_6=\frac{1}{FH^2}\, .
\end{equation}
It is more convenient for the purposes of this paper to use the
$e$-foldings number, so by using the following differentiation rule,
\begin{equation}\label{specialderivative}
\frac{\mathrm{d}}{\mathrm{d}N}=\frac{1}{H}\frac{\mathrm{d}}{\mathrm{d}t}\,
,
\end{equation}
and by taking the first derivative of the variables
(\ref{variablesslowdown}), in conjunction with Eqs.
(\ref{eqnsfggrav}), after some algebra we obtain,
\begin{align}\label{dynamicalsystemmain}
&
\frac{\mathrm{d}x_1}{\mathrm{d}N}=\frac{1}{4}(x_3-1)x_6+8x_1-(x_3-1)x_1+\frac{1}{2}x_4x_6+\frac{3}{8}x_5x_6+x_1^2\,
,
\\ \notag &
\frac{\mathrm{d}x_2}{\mathrm{d}N}=-\frac{16}{x_6}+\frac{8}{x_6}m-\frac{32}{x_6}(x_3-1) \, ,\\
\notag & \frac{\mathrm{d}x_3}{\mathrm{d}N}=2(x_3-1)^2+m+\frac{96}{24}(x_3-1)-4x_3(x_3-1) \, , \\
\notag & \frac{\mathrm{d}x_4}{\mathrm{d}N}=-4x_4-2x_4(x_3-1) \, ,\\
\notag & \frac{\mathrm{d}x_5}{\mathrm{d}N}=-3x_5-2x_5(x_3-1) \, ,\\
\notag & \frac{\mathrm{d}x_6}{\mathrm{d}N}=x_1x_6-2(x_3-1)x_6 \, ,
\end{align}
where the parameter $m$ stands for,
\begin{equation}\label{parameterm}
m=\frac{\ddot{H}}{H^3}\, .
\end{equation}
The only time-dependence ($N$-dependence) of the dynamical system is
contained on the parameter $m$, which we shall assume that it is
constant and particularly equal to zero. As we shall see, this case
describes a quasi de Sitter evolution, so the phase space analysis
will characterize inflationary attractors. Also, since we are
interested on inflationary attractors, we shall assume that the
$e$-foldings number takes values in the range $N=[0,60]$.

A useful quantity that will make clear the physical significance of
the fixed points of the dynamical system, is the effective equation
of state parameter (EoS), which is defined as follows,
\begin{equation}\label{weffoneeqn}
w_{eff}=-1-\frac{2\dot{H}}{3H^2}\, ,
\end{equation}
so by using the definition of the parameter $x_3$ from Eq.
(\ref{variablesslowdown}) and also the definition of the
Gauss-Bonnet invariant from Eq. (\ref{gaussbonnetinvariant}), the
EoS can be written as follows,
\begin{equation}\label{eos1}
w_{eff}=-1-\frac{2}{3} (x_3-1)\, .
\end{equation}
By having the dynamical system of Eq. (\ref{dynamicalsystemmain}) at
hand, and also the EoS (\ref{eos1}), in the next section we shall
investigate the inflationary phase space of $f(\mathcal{G})$
gravity, and we shall analyze numerically the stability of the fixed
points.

\section{Analysis of the $f(G)$ Gravity Inflationary Phase Space}

Now let us focus on the study of the dynamical system
(\ref{dynamicalsystemmain}), and as we already discussed, the
parameter $m$ is the only time-dependent parameter in the dynamical
system. We shall assume that the value of $m$ is zero, in which case
the general form of the scale factor that realizes the $m=0$ case,
has the following form,
\begin{equation}\label{quasidesitter}
a(t)=e^{H_0 t-H_i t^2}\, ,
\end{equation}
which in general describes a de Sitter evolution. Notice that by
fixing $m$, the dynamical system contains parameters which remain
general, so in the rest of this section we shall examine the impact
of the value $m=0$ on the dynamical system.

Before getting into  the details of the analysis, let us briefly
present some standard features of dynamical systems analysis, also
found in Ref. \cite{dynsystemsbook}. Particularly, we shall be
interested in the linearization method and the Hartman-Grobman
theorem. The latter determines the stability of the fixed points,
when this is hyperbolic. Assume that the function $\Phi (t)$
$\epsilon$ $R^n$ is a solution to the following dynamical system,
\begin{equation}\label{ds1}
\frac{\mathrm{d}\Phi}{\mathrm{d}t}=g(\Phi (t))\, ,
\end{equation}
where $g(\Phi (t))$ is a locally Lipschitz continuous map
$g:R^n\rightarrow R^n$. We denote with $\phi_*$, the fixed points of
the dynamical system (\ref{ds1}), and also let $\mathcal{J}(g)$ be
the corresponding Jacobian matrix. The latter is equal to,
\begin{equation}\label{jaconiab}
\mathcal{J}=\sum_i\sum_j\Big{[}\frac{\mathrm{\partial f_i}}{\partial
x_j}\Big{]}\, .
\end{equation}
The Jacobian matrix $\mathcal{J}(g)$ should be calculated at the
fixed points, and the corresponding eigenvalues must satisfy
$\mathrm{Re}(e_i)\neq 0$, in order to have a concrete idea on the
stability of the fixed points. Assume that the spectrum of the
eigenvalues of $\mathcal{J}(g)$, is $\sigma (\mathcal{J}(g))$, then
a hyperbolic fixed point satisfies $\mathrm{Re}\left(\sigma
(\mathcal(J))\right)\neq 0$. The Hartman theorem, when applied for a
autonomous system, indicates the certain existence of a homeomorphic
map $\mathcal{F}:U\rightarrow R^n$, with $U$ being an open
neighborhood of the fixed point $\phi_*$, which satisfies
$\mathcal{F}(\phi_*)$. The flow generated by the homeomorphism , is,
\begin{equation}\label{fklow}
\frac{\mathrm{d}h(u)}{\mathrm{d}t}=\mathcal{J}h(u)\, ,
\end{equation}
which is topologically equivalent to the one appearing in
(\ref{ds1}). By applying the Hartman theorem, the dynamical system
(\ref{ds1}), can be written in the following way,
\begin{equation}\label{dapprox}
\frac{\mathrm{d}\Phi}{\mathrm{d}t}=\mathcal{J}(g)(\Phi)\Big{|}_{\Phi=\phi_*}
(\Phi-\phi_*)+\mathcal{S}(\phi_*,t)\, ,
\end{equation}
with $\mathcal{S}(\phi,t)$ being a smooth map $[0,\infty )\times
R^n$. Hence, if the Jacobian matrix satisfies
$\mathcal{Re}\left(\sigma (\mathcal{J}(g))\right)<0$, and in
addition, if the following holds true,
\begin{equation}\label{gfgd}
\lim_{\Phi\rightarrow
\phi_*}\frac{|\mathcal{S}(\phi,t)|}{|\Phi-\phi_*|}\rightarrow 0\, ,
\end{equation}
the fixed point $\phi_*$ of the dynamical flow
$\frac{\mathrm{d}\Phi}{\mathrm{d}t}=\mathcal{J}(g)(\Phi)\Big{|}_{\Phi=\phi_*
}(\Phi-\phi_*)$, is also a fixed point of the flow (\ref{dapprox}),
and moreover, it is asymptotically stable. Hence, when a hyperbolic
fixed point is met, the above statements hold true, and in the
converse case, further analysis, supported by numerical studies, are
required in order to reveal the stability of the fixed points. This
is in fact our strategy, since the resulting fixed points for the
$f(\mathcal{G})$ gravity, are not hyperbolic.


So let us fix $m=0$, and we proceed to the study of the dynamical
system. For the dynamical system of Eq. (\ref{dynamicalsystemmain}),
the matrix $\mathcal{J}=\sum_i\sum_j\Big{[}\frac{\mathrm{\partial
f_i}}{\partial x_j}\Big{]}$ reads,
\begin{equation}\label{matrixceas}
\mathcal{J}=\left(
\begin{array}{cccccc}
 2 x_1-x_3+9 & 0 & \frac{x_6}{4}-x_1 & \frac{x_6}{2} & \frac{3 x_6}{8} & \frac{x_3-1}{4}+\frac{x_4}{2}+\frac{3 x_5}{8} \\
 0 & -2 (x_3-1) & -2 x_2-\frac{32}{x_6} & 0 & 0 & \frac{32 (x_3-1)}{x_6^2}+\frac{16}{x_6^2} \\
 0 & 0 & 4-4 x_3 & 0 & 0 & 0 \\
 0 & 0 & -2 x_4 & -2 (x_3-1)-4 & 0 & 0 \\
 0 & 0 & -2 x_5 & 0 & -2 (x_3-1)-3 & 0 \\
 x_6 & 0 & -2 x_6 & 0 & 0 & x_1-2 (x_3-1) \\
\end{array}
\right)\, ,
\end{equation}
where in this case, the functions $f_i$ are,
\begin{align}\label{fis}
&
f_1=\frac{1}{4}(x_3-1)x_6+8x_1-(x_3-1)x_1+\frac{1}{2}x_4x_6+\frac{3}{8}x_5x_6\,
, \\ \notag &
f_2=-\frac{16}{x_6}+\frac{8}{x_6}m-\frac{32}{x_6}(x_3-1) ,\\ \notag
& f_3=2(x_3-1)^2+m+\frac{96}{24}(x_3-1)-4x_3(x_3-1),\\ \notag &
f_4=-4x_4-2x_4(x_3-1),\\ \notag & f_5=-3x_5-2x_5(x_3-1),\\ \notag &
f_6=x_1x_6-2(x_3-1)x_6 \, .
\end{align}
We can easily calculate the fixed points of the dynamical system
(\ref{dynamicalsystemmain}), which for general $m$ are,
\begin{align}\label{fixedpointsgeneral}
& \phi_*^1=(-\sqrt{2} \sqrt{m},\frac{6 \sqrt{2} m^{3/2}+m^2-30 m+16
\sqrt{2} \sqrt{m}}{m^2-128 m},\frac{1}{2} \left(2-\sqrt{2}
\sqrt{m}\right),0,0,4 \left(\sqrt{2} \sqrt{m}-16\right))
 \\ \notag & \phi_*^2=(\sqrt{2} \sqrt{m},\frac{-6 \sqrt{2} m^{3/2}+m^2-30 m-16 \sqrt{2} \sqrt{m}}{m^2-128 m},\frac{1}{2} \left(\sqrt{2} \sqrt{m}+2\right),0,0,-4 \left(\sqrt{2} \sqrt{m}+16\right)),\, .
\end{align}
and in the case $m=0$, the fixed points coincide and we thus have
only the following fixed point,
\begin{equation}\label{fixedpointdesitter}
\phi_*^1=(0,\infty,1,0,0,-64)\, .
\end{equation}
The corresponding eigenvalues for the fixed point $\phi_*^1$ are
$(8,-4,-3,0,0,0)$, so it is not hyperbolic for sure, and as we will
demonstrate it is also strongly unstable. Before proceeding to the
numerical analysis, let us reveal the physical significance of the
fixed point $\phi_*^1$, and to this end let us investigate the
behavior of the EoS for the fixed point $\phi_*^1$. So for $x_3=1$,
the EoS (\ref{eos1}) becomes $w_{eff}=-1$, and in effect, the fixed
point is a de Sitter fixed point.

\begin{figure}[h]
\centering
\includegraphics[width=20pc]{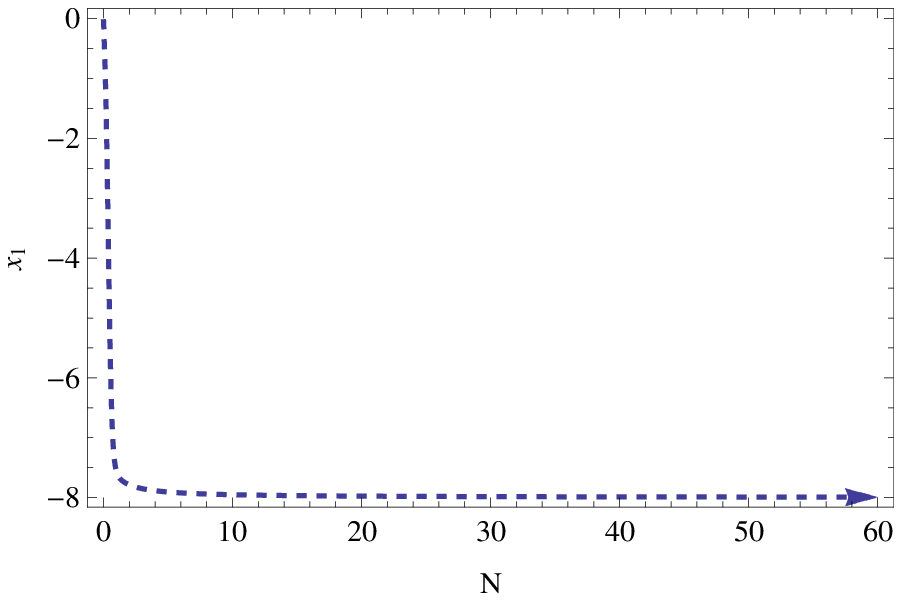}
\includegraphics[width=20pc]{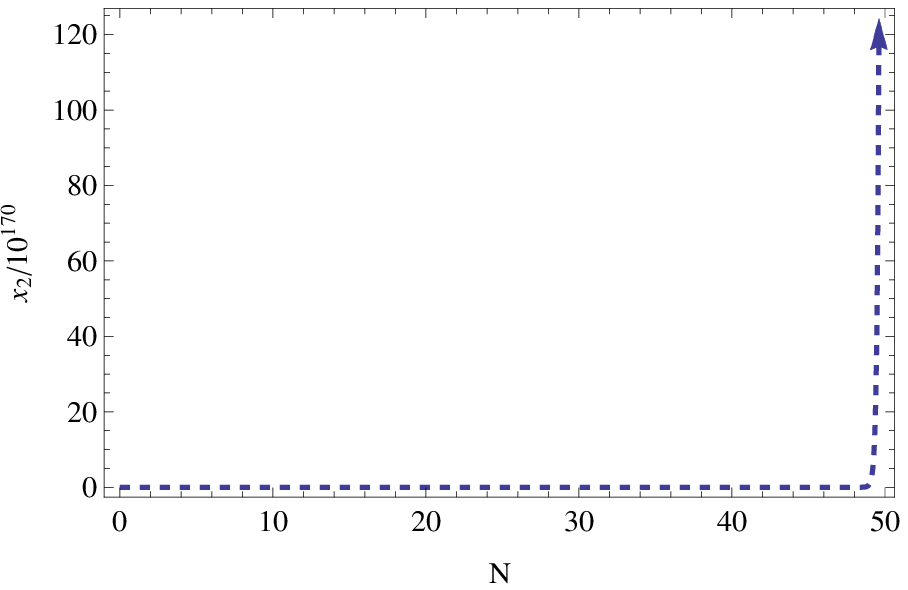}
\includegraphics[width=20pc]{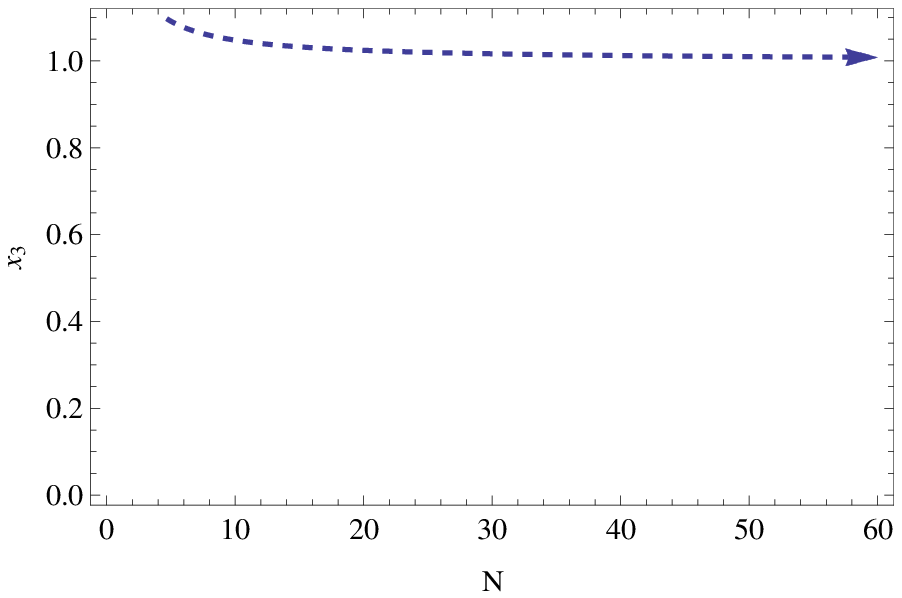}
\caption{{\it{The behavior of $x_1(N)$, $x_2(N)$ and $x_3(N)$ for
the dynamical system (\ref{dynamicalsystemmain}), for the initial
conditions $x_1(0)=-0.01$, $x_2(0)=0$ and $x_3(0)=2.05$, $x_4(0)=0$,
$x_5(0)=7$, $x_6(0)=-2$,  and for $m=0$. }}} \label{plot1}
\end{figure}

\begin{figure}[h]
\centering
\includegraphics[width=20pc]{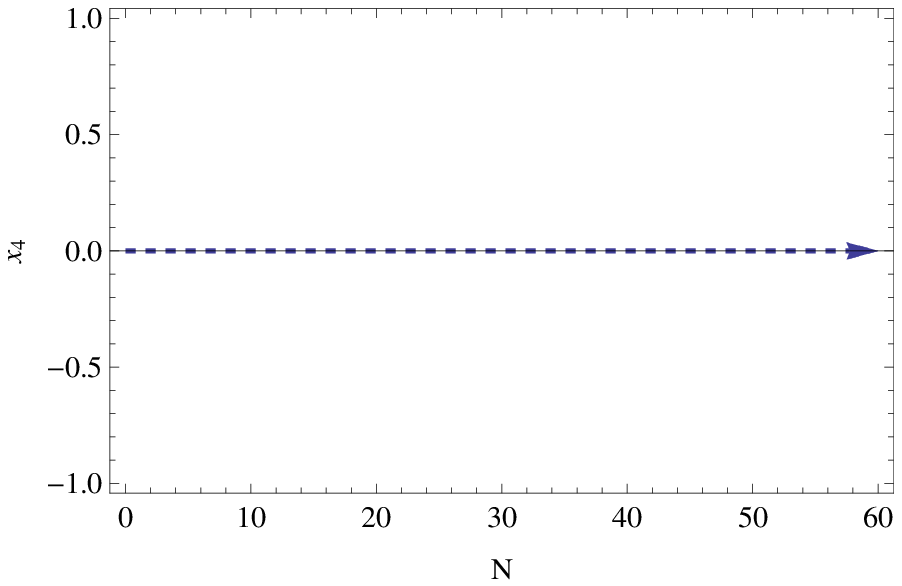}
\includegraphics[width=20pc]{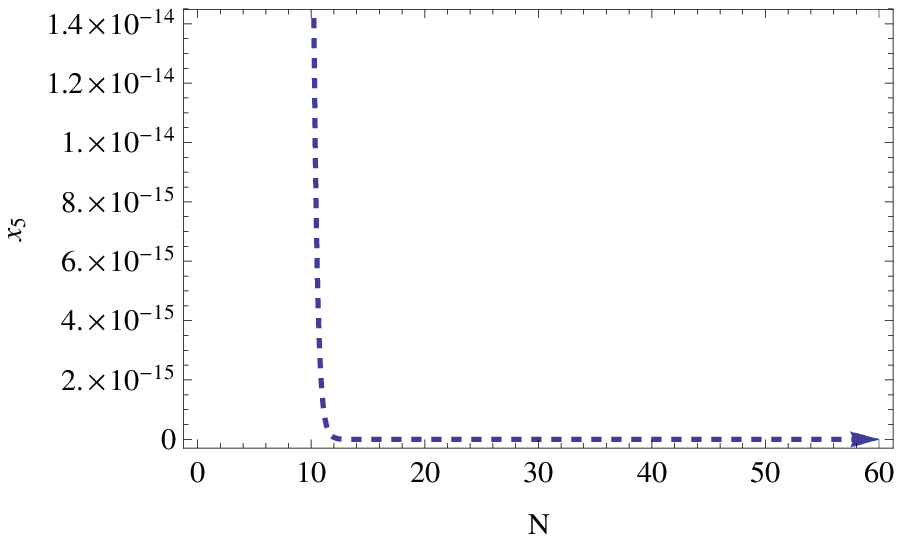}
\includegraphics[width=20pc]{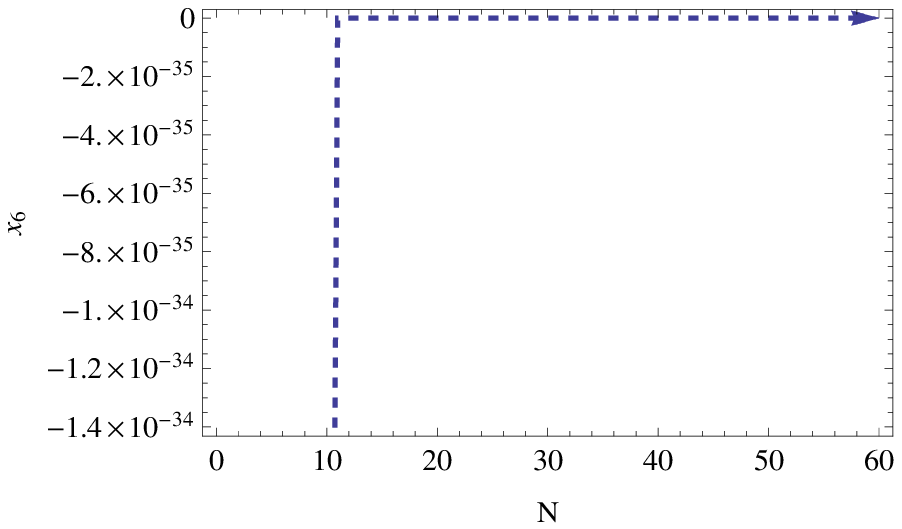}
\caption{{\it{The behavior of $x_4(N)$, $x_5(N)$ and $x_6(N)$ for
the dynamical system (\ref{dynamicalsystemmain}), for the initial
conditions $x_1(0)=-0.01$, $x_2(0)=0$ and $x_3(0)=2.05$, $x_4(0)=0$,
$x_5(0)=7$, $x_6(0)=-2$,  and for $m=0$. }}} \label{plot2}
\end{figure}

The instability of the fixed point can be revealed only numerically,
since the Hartman-Grobman theorem does not apply in our case, due to
the fact that the fixed point is not hyperbolic, so by solving
numerically the dynamical system (\ref{dynamicalsystemmain}) for
various initial conditions, we can conclude whether the de Sitter
fixed point is stable or not. We emphasize our analysis for values
of the $e$-foldings in the range $N=(0,60)$, and in Figs.
\ref{plot1} and \ref{plot2} we plot the numerical solutions for the
dynamical system (\ref{dynamicalsystemmain}),by using the initial
conditions $x_1(0)=-0.01$, $x_2(0)=0$ and $x_3(0)=2.05$, $x_4(0)=0$,
$x_5(0)=7$, $x_6(0)=-2$.
As it can be seen in Fig. \ref{plot1}, the variable $x_1$ tends to
$x_1\to -8$ quite fast, and also $x_2\to \infty$, while $x_3\to 1$,
so the de Sitter fixed point is reached, when the parameter $x_3$ is
taken into account, while $x_1$ does not converge to zero. From Fig.
\ref{plot2}, it can be seen that the variables $x_4$ and $x_5$ tend
to their de Sitter values, however the variable $x_6$ tends to zero
around $N\sim 60$. Hence it is obvious that the de Sitter fixed
point $\phi_*^1$ is not eventually reached from all the variables,
and therefore it is an unstable fixed point.
Now let us turn our focus on the purely vacuum case, for which
$\rho_m=\rho_r=0$. Actually, the fixed point $\phi_*^1$ has features
of a vacuum fixed point since $x_4=x_5=0$, but we shall investigate
separately the two cases for clarity.

\begin{figure}[h]
\centering
\includegraphics[width=20pc]{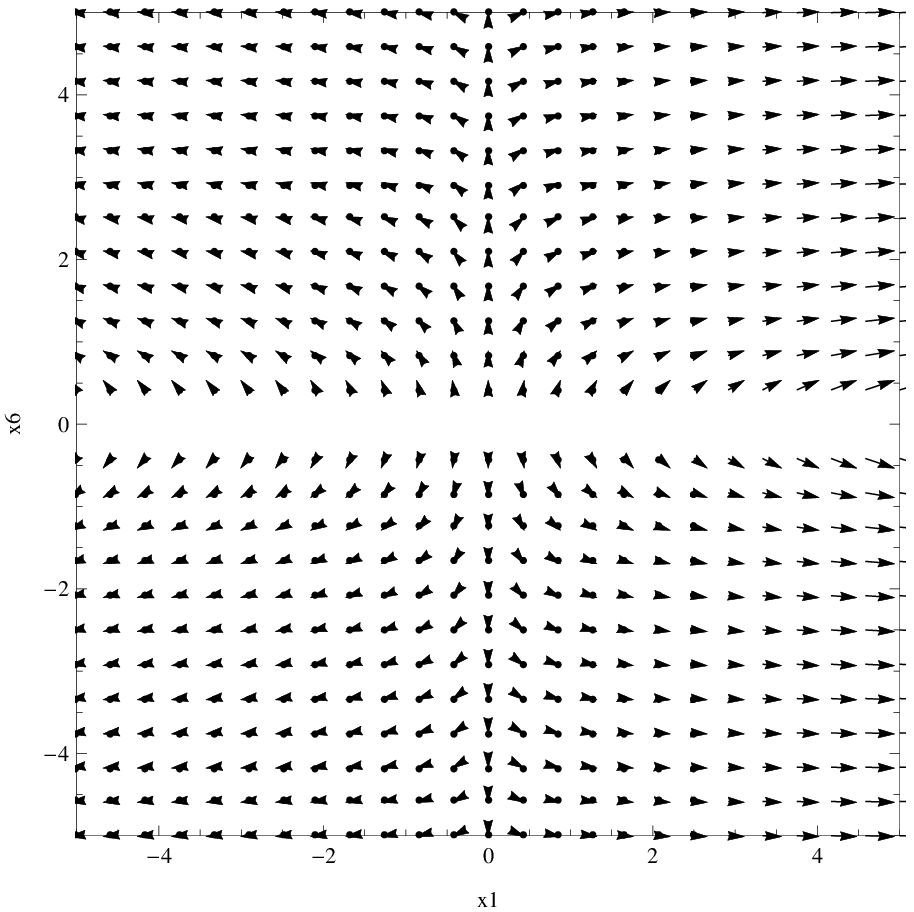}
\includegraphics[width=20pc]{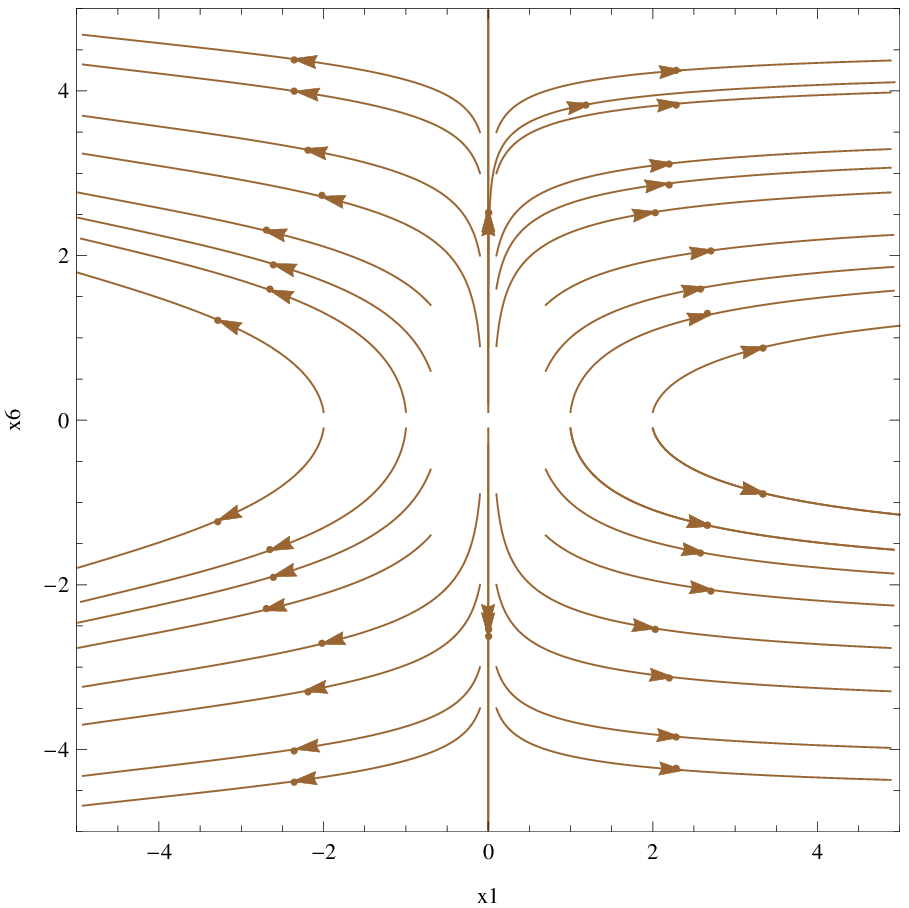}
\caption{{\it{2-dimensional flow (left) and various trajectories in
the $x_1-x_6$ plane (right) for the dynamical system
(\ref{dynamicalsystemmain12}), and for $m=0$.}}} \label{plot3}
\end{figure}

In the case that matter and radiation are excluded, the dynamical
system (\ref{dynamicalsystemmain}), takes the following form,
\begin{align}\label{dynamicalsystemmain1}
&
\frac{\mathrm{d}x_1}{\mathrm{d}N}=\frac{1}{4}(x_3-1)x_6+8x_1-(x_3-1)x_1+x_1^2\,
,
\\ \notag &
\frac{\mathrm{d}x_2}{\mathrm{d}N}=-\frac{16}{x_6}+\frac{8}{x_6}m-\frac{32}{x_6}(x_3-1) \, ,\\
\notag & \frac{\mathrm{d}x_3}{\mathrm{d}N}=2(x_3-1)^2+m+\frac{96}{24}(x_3-1)-4x_3(x_3-1) \, , \\
\notag & \frac{\mathrm{d}x_6}{\mathrm{d}N}=x_1x_6-2(x_3-1)x_6 \, ,
\end{align}
and the corresponding matrix
$\mathcal{J}=\sum_i\sum_j\Big{[}\frac{\mathrm{\partial
f_i}}{\partial x_j}\Big{]}$ becomes in this case,
\begin{equation}\label{matrixceas1}
\mathcal{J}=\left(
\begin{array}{cccc}
 2 x_1-x_3+9 & 0 & \frac{x_6}{4}-x_1 & \frac{x_3-1}{4} \\
 0 & -2 (x_3-1) & -2 x_2-\frac{32}{x_6} & \frac{8 m}{x_6^2}+\frac{32 (x_3-1)}{x_6^2}+\frac{16}{x_6^2} \\
 0 & 0 & 4-4 x_3 & 0 \\
 x_6 & 0 & -2 x_6 & x_1-2 (x_3-1) \\
\end{array}
\right)\, ,
\end{equation}
and in this case, the functions $f_i$ are,
\begin{align}\label{fis1}
& f_1=\frac{1}{4}(x_3-1)x_6+8x_1-(x_3-1)x_1\, , \\
\notag & f_2=-\frac{16}{x_6}+\frac{8}{x_6}m-\frac{32}{x_6}(x_3-1)
,\\ \notag & f_3=2(x_3-1)^2+m+\frac{96}{24}(x_3-1)-4x_3(x_3-1),\\
\notag & f_6=x_1x_6-2(x_3-1)x_6 \, .
\end{align}
The corresponding fixed point in the $m=0$ case is as expected,
\begin{equation}\label{fixedpointdesitter1}
\varphi_*^1=(0,\infty,1,-64)\, .
\end{equation}
From a physical point of view, the dynamical evolution is
qualitatively the same with the scenario we described in the
non-vacuum case, due to the fact that $x_4=x_5=0$. This can be
verified by a numerical analysis which we omit for brevity. Hence,
the de Sitter fixed point exists in this case too, and it is
unstable. In order to further support this result, we shall present
another aspect of our numerical analysis, emphasizing in the
dynamical system which corresponds to $x_3=1$. In this case, the
dynamical system becomes,
\begin{align}\label{dynamicalsystemmain12}
& \frac{\mathrm{d}x_1}{\mathrm{d}N}=8x_1+x_1^2\, ,
\\ \notag &
\frac{\mathrm{d}x_2}{\mathrm{d}N}=-\frac{16}{x_6}+\frac{8}{x_6}m \, , \\
\notag & \frac{\mathrm{d}x_6}{\mathrm{d}N}=x_1x_6 \, ,
\end{align}
and in Fig. \ref{plot3}, we present the behavior of the reduced
system in terms of the variables $(x_1,x_6)$. In the left plot, the
vector field flow appears, in the $x_1-x_6$ plane, while in the
right plot, the behavior of various trajectories appear in the
$x_1-x_6$ plane. As it can be seen in this case too, the reduced
system is unstable at $x_1=0$, which is the value of the variable
$x_1$ at the fixed point $\varphi_*^1$ in Eq.
(\ref{fixedpointdesitter1}).
Hence, the autonomous dynamical system (\ref{dynamicalsystemmain})
both in the vacuum and in the presence of radiation and matter, has
a de Sitter fixed point, which is unstable. The instability is
mainly caused by the parameter $x_1$, since it never reaches its
fixed point value $x_1=0$, and as it seen by the plots in Fig.
\ref{plot3}, the dynamical system trajectories and flow are repelled
away from the value $x_1=0$. This instability could be an indicator
of a graceful exit from inflation mechanism inherent in the
$f(\mathcal{G})$ gravity, both in the vacuum and non-vacuum cases,
however a closer analysis on this is required, which we hope to
address in a future work.

\section{Conclusions}

In this paper we performed a detailed numerical analysis of the
$f(\mathcal{G})$ gravity phase space, in the case that the
corresponding dynamical system can be formed to be an autonomous
dynamical system. As we demonstrated, by appropriately choosing the
free independent variables, the dynamical system can be an
autonomous dynamical system, with the only time-dependence being
contained in the parameter $m$. We focused on the case $m=0$, which
describes a quasi-de Sitter evolution in the most general case, and
we investigated how the dynamical evolution behaves in this case. As
we demonstrated, the dynamical system has a de Sitter fixed point,
for which the EoS is $w_{eff}=-1$, and we examined the behavior of
the variables numerically. The resulting picture indicated that the
fixed point is unstable, a feature that could possibly indicate that
the $f(\mathcal{G})$ gravity has an inherent instability mechanism,
traced on the instability of the variable $x_1$, which may be seen
as a graceful exit from inflation mechanism. This feature however,
needs closer inspection, which we plan to do in a future work. Finally, it would be interesting to combine the dynamical system study with the Noether symmetry approach, see for example \cite{Capozziello:2014ioa} for a Gauss-Bonnet theory account.


\end{document}